\documentclass[12pt]{article}
\usepackage{latexsym,epsfig,graphicx,a4wide,amssymb}

{\rm }

\def\be{\begin{equation}}
 \def\ee{\end{equation}}
 \def\bea{\begin{eqnarray}}
 \def\eea{\end{eqnarray}}

\newcommand{\fr}{\frac}
\newcommand{\pr}{\prime}

\def\2{\frac{1}{2}}
\def\4{\frac{1}{4}}

\def\gen{\mathrm{g}}

\catcode`\@=11

\@addtoreset{equation}{section}

\def\@normalsize{\@setsize\normalsize{15pt}\xiipt\@xiipt
\abovedisplayskip 14pt plus3pt minus3pt%
\belowdisplayskip \abovedisplayskip
\abovedisplayshortskip  \z@ plus3pt%
\belowdisplayshortskip  7pt plus3.5pt minus0pt}
\def\small{\@setsize\small{13.6pt}\xipt\@xipt
\abovedisplayskip 13pt plus3pt minus3pt%
\belowdisplayskip \abovedisplayskip
\abovedisplayshortskip  \z@ plus3pt%
\belowdisplayshortskip  7pt plus3.5pt minus0pt
\def\@listi{\parsep 4.5pt plus 2pt minus 1pt
            \itemsep \parsep
            \topsep 9pt plus 3pt minus 3pt}}

\def\underline#1{\relax\ifmmode\@@underline#1\else
        $\@@underline{\hbox{#1}}$\relax\fi}
\@twosidetrue \relax

\catcode`@=12

\catcode`\@=11

\def\section{\@startsection{section}{1}{\z@}{3.5ex plus 1ex minus
   .2ex}{2.3ex plus .2ex}{\large\bf}}

\def\ps@headings{\def\@oddfoot{}\def\@evenfoot{}
\def\@oddhead{\hbox{}\hfill
        \makebox[.5\textwidth]{\raggedright\ignorespaces --\thepage{}--
        \hfill }}
\def\@evenhead{\@oddhead}
\def\subsectionmark##1{\markboth{##1}{}}
}

\ps@headings

\catcode`\@=12

\begin{document}

\begin{titlepage}

\rightline{UTHET-09-0102}

\begin{centering}
\vspace{1cm}
{\Large {\bf Exact Gravity Dual of a Gapless Superconductor
}}\\

\vspace{1.5cm}

 {\bf George Koutsoumbas $^{\sharp}$}, {\bf Eleftherios Papantonopoulos $^{*}$} \\
 \vspace{.2in}
 Department of Physics, National Technical University of
Athens, \\
Zografou Campus GR 157 73, Athens, Greece \\
\vspace{.2in}
 {\bf George Siopsis $^{\flat}$}
\vspace{.2in}

 Department of Physics and Astronomy, The
University of Tennessee,\\ Knoxville, TN 37996 - 1200, USA
 \\
\vspace{3mm}

\end{centering}
\vspace{2cm}

\begin{abstract}

A model of an exact gravity dual of a gapless superconductor is presented
in which the condensate is provided by a charged scalar field
coupled to a bulk black hole of hyperbolic horizon in asymptotically AdS spacetime.
A critical temperature exists at which the mass of the black hole vanishes and
a scaling symmetry emerges.
 Below the critical point, the black hole acquires its hair through  a  phase transition while   an electromagnetic perturbation of
the background Maxwell field determines the conductivity
of the boundary theory.

\end{abstract}

\vspace{3.5cm}
\begin{flushleft}
$^{\sharp}~~$ e-mail address: kutsubas@central.ntua.gr \\
$^{*} ~~$ e-mail address: lpapa@central.ntua.gr \\
$ ^{\flat}~~$ e-mail address: siopsis@tennessee.edu

\end{flushleft}
\end{titlepage}

\section{Introduction}

The AdS/CFT correspondence has become a powerful tool in studying
strongly coupled phenomena in quantum field theory using results
from a weakly coupled gravity background. According to this
correspondence principle~\cite{Maldacena:1997re}, a string theory
on asymptotically AdS spacetimes can be related to a conformal
field theory on the boundary. In recent years, apart from string
theory, this holographic correspondence, following a more
phenomenological approach, has also been applied to nuclear
physics in order to describe certain aspects such as heavy ion
collisions at RHIC \cite{Mateos:2007ay} and to certain condensed
matter systems. Phenomena such as the Hall effect
\cite{Hartnoll:2007ai} and Nernst effect \cite{Hartnoll:2007ih,
Hartnoll:2007ip, Hartnoll:2008hs} have dual gravitational
descriptions.

Recently the AdS/CFT correspondence has also been applied to
superconductivity~\cite{Hartnoll:2008vx}. The gravity dual of a
superconductor consists of a system with a black hole and a
charged scalar field, in which the black hole admits scalar hair
at temperature smaller than a critical temperature, while there is
no scalar hair at larger temperatures. A condensate of the charged
scalar field is formed through its coupling to a Maxwell field of
the background. Neither field was backreacting on the metric. This
decoupled Abelian-Higgs sector can be obtained from an
Einstein-Maxwell-scalar theory~\cite{Gubser:2008px} through a
scaling limit in which the product of the charge of the black hole
and the charge of the scalar field is held fixed while the latter
is taken to infinity. Considering fluctuations of the vector
potential, the frequency dependent conductivity was calculated,
and it was shown that it develops a gap determined by the
condensate.

The model of the gravitational dual to the superconductor
in~\cite{Hartnoll:2008vx} was further studied beyond the probe
limit~\cite{Hartnoll:2008kx}. Away from the large charge limit, the
backreaction of the scalar field to the spacetime metric has to be
taken into consideration. It was found that all the essential
characteristics of the dual superconductor were persisting.
Moreover, even for very small charge the superconductivity was
maintained. These models however are phenomenological models. The
classical fields and their interactions are chosen by hand. It would
have also been desirable that these models emerge from a consistent
string theory~\cite{Ammon:2008fc}.

In this work we propose a model of a gravity dual of a gapless
superconductor in which a charged scalar field provides
the scalar hair of an exact black hole
solution~\cite{Martinez:2004nb,Martinez:2005di}. It has been shown
in~\cite{Koutsoumbas:2006xj} that, below a critical temperature,
this black hole solution undergoes a spontaneous dressing up with
the scalar hair, while above that critical temperature the dressed
black hole
  decays into the bare black hole. At the critical point, the mass of the
  black hole vanishes and a scaling symmetry emerges, because the metric becomes purely AdS.
  We will show that, if the scalar field coupled to gravity in the bulk is charged, a condensate forms, while
  an electromagnetic perturbation  of the background
determines the conductivity and therefore the superfluid density of the
boundary theory. There is evidence that these black hole solutions can be
obtained from eleven-dimensional supergravity
theory~\cite{Papadimitriou:2006dr}.

The paper is organized as follows.
In section \ref{sec2} we discuss an exact black hole
solution with scalar hair and in
 section \ref{sec3} we explain how the black hole
acquires its hair through a third order phase transition. In section
\ref{secst} we outline a stability analysis of hairy black holes. In
section \ref{sec4} we discuss the dual superconductor on the
boundary of the exact hairy black hole solution and calculate its
conductivity and superfluid density analytically. In section
\ref{sec5} we support our analytical results of section \ref{sec4}
with a numerical investigation and present evidence showing that the
superconductor is gapless. Finally, section \ref{sec6} contains our
concluding remarks.

\section{Black Hole with Scalar Hair}
\label{sec2}

To obtain a black hole with scalar hair, we start with the four-dimensional action
\begin{equation}\label{holaction1}
I = I_{gr} + I_{\mathrm{matter}}
\end{equation}
consisting of the Einstein-Hilbert action with a negative cosmological constant $\Lambda = - \frac{3}{l^2}$,
\begin{equation}
I_{gr} = \frac{1}{16\pi G} \int d^4x\sqrt{-g}\left[ R+ \frac{6}{l^2} \right] \end{equation}
where $G$ is Newton's constant, $R$ is the Ricci scalar, $l$ is the AdS radius and a charged scalar together with a Maxwell field (matter fields)
\begin{equation}
I_{\mathrm{matter}} = \int d^4x\sqrt{-g}\left[ g^{\mu\nu} D_\mu\phi (D_\nu\phi)^{*}
-\frac{1}{6}R\phi^*\phi
-\lambda (\phi^* \phi)^{2}\right]
-\frac{1}{4 }\int d^{4}x\sqrt{-g}F^{\mu
\nu }F_{\mu \nu }~,
\end{equation}
where
\begin{equation} D_\mu \phi \equiv \partial_\mu\phi + iqA_\mu \phi \end{equation}
and $\lambda$ is an arbitrary coupling constant. The
corresponding field equations are
\begin{eqnarray}
G_{\mu\nu} -\frac{3}{l^2} g_{\mu\nu}&=&8\pi G T_{\mu\nu}^{\mathrm{matter}}~, \nonumber\\
D_\mu D^\mu \phi&=&\frac{1}{6}R\phi+\lambda\phi |\phi|^2 ~, \nonumber\\
\frac{1}{\sqrt{-g}} \partial _{\nu }(\sqrt{-g}F^{\mu \nu }) &=& J^\mu~,
\end{eqnarray}
where the energy-momentum tensor is given by
\begin{equation}
\label{Tuvfield}T^{\mu\nu}_{\mathrm{matter}}= \frac{1}{\sqrt{-g}} \frac{\delta I_{\mathrm{matter}} }{\delta g_{\mu\nu}}~,
\end{equation}
and the electric current is
\begin{equation}
\label{Tuvem}J_\mu = iq (\phi^* D_\mu \phi - \mathrm{c.c.})
~.
\end{equation}
The presence of
a negative cosmological constant allows the existence of black holes
with topology $\mathbb{R}^{2}\times\Sigma$, where $\Sigma$ is a
two-dimensional manifold of constant negative curvature. These
black holes are known as topological black holes
(TBHs)~\cite{topological, Vanzo:1997gw}. The simplest solution reads
\begin{equation}\label{linel}
ds^{2}=-f_{TBH}(\rho)dt^{2}+\frac{1}{f_{TBH}(\rho)}d\rho^{2}+\rho^{2}d\sigma ^{2}\quad
,\quad f_{TBH}(\rho)=\frac{\rho^{2}}{l^2} -1-\frac{\rho_0 }{\rho}~,
\end{equation}
with $\phi = 0$, $A_\mu = 0$.
$\rho_0$ is a
constant which is proportional to the mass and is bounded from
below ($\rho_0\geq-\frac{2}{3\sqrt{3}} l$).
$d\sigma^{2}$ is the line
element of the two-dimensional manifold $\Sigma$, which is locally
isomorphic to the hyperbolic manifold $H^{2}$ and of the form
\begin{equation}\label{eqSigma}
\Sigma=H^{2}/\Gamma \quad \textrm{,\quad  $\Gamma\subset
O(2,1)$}~,
\end{equation}
where $\Gamma$ is a freely acting discrete subgroup (i.e., without
fixed points) of isometries. The line element $d\sigma^{2}$ of
$\Sigma$ can be written as
\begin{equation}
d\sigma^{2}=d\theta^{2}+\sinh^{2}\theta d\varphi{^2}~,
\end{equation}
with $\theta\ge0$ and $0\le\phi<2\pi$ being the coordinates of the
hyperbolic space $H^{2}$ or pseudosphere, which is a non-compact
two-dimensional space of constant negative curvature. This space
becomes a compact space of constant negative curvature with genus
$\gen\ge2$ by identifying, according to the connection rules of
the discrete subgroup $\Gamma$, the opposite edges of a
$4\gen$-sided polygon whose sides are geodesics and is centered at
the origin $\theta=\varphi=0$ of the
pseudosphere~\cite{topological, Vanzo:1997gw, Balazs:1986uj}. An
octagon is the simplest such polygon, yielding a compact surface
of genus $\gen=2$ under these identifications. Thus, the
two-dimensional manifold $\Sigma$ is a compact Riemann 2-surface
of genus $\mathrm{g}\geq2$.  The configuration (\ref{linel}) is an
asymptotically locally AdS spacetime. The horizon structure of
(\ref{linel}) is determined by the roots of the metric function
$f_{TBH}(\rho)$, that is
\begin{equation}
f_{TBH}(\rho)=\frac{\rho^{2}}{l^2}-1-\frac{\rho_0}{\rho}=0~.
\end{equation}
For $-\frac{2}{3\sqrt{3}} l <\rho_0<0$, this equation has two distinct
non-degenerate solutions, corresponding to an inner and an outer
horizon $\rho_{-}$ and $\rho_{+}$ respectively. For $\rho_0\geq0$, $f_{TBH}(\rho)$
has just one non-degenerate root and so the black hole
(\ref{linel}) has one horizon $\rho_{+}$. The horizons for both cases
of $\rho_0$ have the non-trivial topology of the manifold $\Sigma$.
We note that for $\rho_0=-\frac{2}{3\sqrt{3}} l$, $f_{TBH}(\rho)$ has a
degenerate root, but this horizon does not have an interpretation
as black hole horizon~\cite{topological}.

The boundary has metric
\be\label{eqnew} ds_\partial^2 = -dt^2 + l^2 d\sigma^2 \ee
so spatially it is a hyperbolic manifold of radius $l$ (and of curvature $-1/l$).

The temperature, entropy and mass of the black hole are,
respectively, \be T=\frac{3}{4 \pi l} \left( \frac{\rho_{+}}{l} - \frac{l}{3\rho_+} \right)~, \quad
S_{TBH}=\frac{\sigma \rho^{2}_{+}}{4G}~,\quad
M_{TBH}=\frac{\sigma\rho_{+}}{8 \pi
G} \left( \frac{\rho_+^2}{l^2} - 1 \right)~.\label{relations2} \ee obeying the law of thermodynamics
$dM=TdS$.

A static black hole solution with topology $\mathbb{R}^{2}\times
\Sigma $ and scalar hair, is given by
 (MTZ black hole)~\cite{Martinez:2004nb}
\begin{equation}\label{MTZconf}
d{s}^{2}=-f_{MTZ} (r) dt^{2}+
\frac{dr^{2}}{f_{MTZ} (r)} +r^{2}d\sigma^{2} \quad, \quad f_{MTZ} = \frac{r^{2}}{l^2}-\biggl(1+\frac{r_0}{r}\biggr)^{2}~,
\end{equation}
with
\begin{equation}\label{Psiconf}
\phi(r)= \frac{1}{\sqrt 2}\Psi(r)~~,~~~~ A_\mu = 0 ~.
\end{equation}
where \be\label{Psiconf1} \Psi (r) \equiv -\sqrt{\frac{3}{4\pi
G}}\frac{r_0}{r+r_0} \ee is the form of the scalar hair found in the
case of a {\it real} scalar
field~\cite{Martinez:2004nb,Martinez:2005di}. Our normalization is
slightly different due to the fact that we have a {\it complex}
scalar field.  Also, we have chosen a negative sign in
(\ref{Psiconf1}) so that the condensates take on positive values.

The above hairy black hole solution is obtained for the special value of the coupling constant
\begin{equation} \lambda = \frac{8\pi G}{3l^2}~. \end{equation}
Other solutions (charged black holes) also exist for values of
$\lambda$ below the above  critical value in the case of a {\it
real} scalar field~\cite{Martinez:2005di}. It would be interesting
to extend those solutions to the case of a {\it complex} scalar
field $\phi$ as well.

The conformally coupled scalar field $|\phi|$ can be turned into a
minimally coupled field through the field redefinition \be \hat
g_{\mu\nu} = \left( 1 - \frac{8\pi G}{3} |\phi|^2 \right)
g_{\mu\nu} \ \ , \ \ \ \ \hat\phi = \sqrt{\frac{3}{8\pi G}}
\tanh^{-1} \sqrt{\frac{8\pi G}{3}} |\phi|~. \ee The action
involving the real scalar field $\hat\phi$ is \be\label{eqxI} I_{\hat\phi} =
\int d^4 x \sqrt{-\hat g} \left[ \frac{1}{2} \hat
g^{\mu\nu}\partial_\mu \hat\phi \partial_\nu \hat\phi -
V(\hat\phi)
+ \dots \right] \ee where
\begin{equation}\label{eqxV}
V(\hat\phi)=-\frac{3}{4 \pi Gl^2}\sinh^{2}\sqrt{\frac{4 \pi G}{3}}\hat\phi~.
\end{equation}
and the dots
represent interaction terms involving other fields, showing that the mass of the scalar
field is $m^2 = -2/l^2$. This mass satisfies the
Breitenlohner-Friedman (BF) bound ensuring the perturbative
stability of AdS spacetime~\cite{Breitenlohner:1982jf,M-T}.
However, the BF bound does not guarantee
the nonlinear stability of
hairy black holes
under general
boundary conditions and potentials.
Therefore, a careful treatment of the stability
issue is needed. In section \ref{secst} we give the main results of the
stability analysis while more details can be found
in~\cite{future}.

The temperature, entropy and mass of the black hole are given
respectively by~\cite{Martinez:2004nb} \be T=\frac{1}{
\pi} \left( \frac{r_+}{l} - \frac{1}{2} \right)~,\quad S_{MTZ}=\frac{\sigma l^2}{4G} \left( 2\frac{r_{+}}{l}-1 \right)~,\quad
M_{MTZ}=\frac{\sigma r_{+}}{4 \pi
G} \left( \frac{r_+}{l} - 1 \right)~.\label{relations1} \ee  It is easy to show that the law of
thermodynamics $dM=TdS$ holds.

For non-negative mass, $M_{MTZ} \geq{0}$, this black hole possesses only
one event horizon at
\begin{equation}
r_{+}=\frac{l}{2}\left(1+\sqrt{1+4\frac{r_0}{l}}\right)~,
\end{equation}
and $\phi$ is regular everywhere. For negative masses,
$-l/4<r_0<0$, the metric (\ref{MTZconf}) develops three horizons,
two of which are event horizons located at $r_{--}$ and at $r_{+}$
 \bea
r_{--}&=&\frac{l}{2}\left(-1+\sqrt{1-4\frac{r_0}{l}}\right)~,
\nonumber \\
 r_{-}&=&\frac{l}{2}\left(1-\sqrt{1+4\frac{r_0}{l}}\right)~,
 \nonumber \\
r_{+}&=&\frac{l}{2}\left(1+\sqrt{1+4\frac{r_0}{l}}\right)~,\label{horizon}
 \eea
 \\
which satisfy $0<r_{--}<-r_0<r_{-}<l/2<r_{+}$. The scalar field
$\phi$ is singular at $r=-r_0$.

Note that if
$\rho_0=0$, $r_0=0$, then both the MTZ  black hole (\ref{MTZconf}) and the TBH
black hole (\ref{linel}) go to
\begin{equation}
ds_{\mathrm{AdS}}^{2}=-\left[ \frac{r^{2}}{l^2} -1\right] dt^{2}+\left[ \frac{r^{2}}{l^2}-1\right]
^{-1}dr ^{2}+r ^{2}d\sigma ^{2}\;, \label{muzero}\end{equation}
which is a
manifold of negative constant curvature possessing an event
horizon at $r=l$. The
MTZ and TBH black holes match continuously at the critical temperature
\be\label{eqTcr} T_0 = \frac{1}{2\pi l} \ee
which corresponds to $M_{TBH} = M_{MTZ} = 0$, with
(\ref{muzero}) being a transient configuration.
Evidently, at the critical point (\ref{eqTcr}) a scaling symmetry emerges owing to the fact that the metric becomes pure AdS.

\section{Phase Transition}
\label{sec3}

In this section we will review the results discussed
in~\cite{Koutsoumbas:2006xj} of a phase transition of a vacuum TBH
to MTZ below the critical temperature (\ref{eqTcr}).

Defining the free energy as
$F=M-TS$ and using relations (\ref{relations1}), we obtain \be
F_{MTZ}=-\frac{\sigma l}{8\pi G}\left( \frac{2r^{2}_{+}}{l^2}-\frac{2r_{+}}{l}+1 \right)~.
\label{free-energy}\ee The free energy~(\ref{free-energy}) can be
written in terms of the temperature as \be F_{MTZ}=-\frac{\sigma l}{8\pi
G}\Big{(}1+2\pi l (T-T_{0}) +2\pi^2 l^2 (T-T_{0})^{2}
\Big{)}~,\label{mtzexpan}\ee where $T_{0}\approx 0.160/l$ is
the critical temperature (\ref{eqTcr}).

On the other hand, the free energy of the TBH, using relations
(\ref{relations2}), is \be F_{TBH}=-\frac{\sigma \rho_+}{16\pi G}
\left( \frac{\rho^{2}_{+}}{l^2}+ 1 \right)~, \label{FETBH} \ee which
can be expanded around the critical temperature $T_{0}$ as \be
F_{TBH}=-\frac{\sigma l}{8\pi G}\Big{(}1+2\pi l (T-T_{0}) +2\pi^2
l^2 (T-T_{0})^{2} + \pi^3 l^3 (T-T_{0})^{3}+\dots
\Big{)}~.\label{tbhexpan}\ee Using (\ref{mtzexpan}) and
(\ref{tbhexpan}), we can calculate the difference between the  TBH
and MTZ  free energies. We obtain \be \Delta F=
F_{TBH}-F_{MTZ}=-\frac{\sigma l}{8\pi G}\ \pi^3 l^3
(T-T_{0})^{3}+\dots~,\label{difference} \ee indicating a third order
phase transition between MTZ and TBH at the critical temperature
$T_0$. Matching the temperatures of the MTZ black hole and the TBH
we obtain \be   \frac{r_+}{l} = \frac{3 \rho_+}{4l}-\frac{l}{4
\rho_+}+\frac{1}{2}~.\label{rhor} \ee It is easily seen that $r_+
\le \rho_+,$ and the inequality is saturated for $r_+ = \rho_+=l.$
We remark that the temperature $T$ should be non-negative, so $r_+
\ge \fr{l}{2}$ for the MTZ black hole and $\rho_+ \ge
\fr{l}{\sqrt{3}}$ for the TBH.

Thermodynamically we can understand this phase transition as
follows. Using relations (\ref{relations1}), (\ref{relations2})
and (\ref{rhor}), we find that $S_{TBH}>S_{MTZ}$ and
$M_{TBH}>M_{MTZ}$ for the relevant ranges of the horizons $r_+$ or
$\rho_+.$ If $r_{+}>l$~ (i.e., the radius of the horizon exceeds the radius of the boundary and $T>T_{0}$), both black holes have
positive mass. As $T>T_0$ implies $F_{TBH} \le F_{MTZ},$ the MTZ
black hole dressed with the scalar field will decay into the bare
black hole. In the decay process, the scalar black hole absorbs
energy from the thermal bath, increasing its horizon radius (from
$r_+$ to $\rho_+>r_+$) and consequently its entropy. Therefore, in
a sense the scalar field is absorbed by the black hole.

If $r_{+}<l$~ (i.e., the radius of the boundary exceeds the radius of the horizon and $T<T_{0}$), both black holes have negative mass,
but now $F_{TBH} > F_{MTZ},$ which means that the MTZ
configuration with nonzero scalar field is favorable.  As a
consequence, below the critical temperature, the bare black hole
undergoes a spontaneous ``dressing up'' with the scalar field. In
the process, the mass and entropy of the black hole decrease and
the differences in energy and entropy are transferred to the heat
bath.

At the critical temperature, the thermodynamic functions of the two
phases match continuously, hence the phase transition is not of
first order. The order parameter that characterizes the transition
can be defined in terms of the value of the scalar field at the
horizon; using the solution for the scalar field (\ref{Psiconf}) we
obtain for $T<T_0$, \be\lambda_{\phi} =\left| \tanh \sqrt{\frac{8\pi
G}{3}}\phi (r_{+})\right|~= \left|\frac{r_{+}-l}{r_{+}}\right| =
\frac{T_{0}-T}{T_{0}+T} \ \ . \ee For $T>T_0,$ $\lambda_\phi$
vanishes. Therefore the order parameter $\lambda_\phi (T)$ is
continuous at the  critical temperature but its first derivative
jumps from $-\frac{1}{2T_0}$ to $0$ as we cross the critical point.


The pure AdS space of (\ref{muzero}) has free energy
$F_{AdS}=-\frac{\sigma l}{8\pi G}$ as can be easily seen using relations
(\ref{free-energy}) or (\ref{FETBH}) with $r_{+}=l$. Then observe
that $F_{AdS}$ is the constant term of both $F_{MTZ}$ in
(\ref{mtzexpan}) and $F_{TBH}$ in (\ref{tbhexpan}). Hence  the
difference of free energies of TBH or MTZ black holes  with the
free energy of pure AdS space indicates that the configuration
(\ref{muzero}) is transient between the MTZ and TBH phase
transition.
At the critical point the mass of the black hole vanishes and a scaling symmetry emerges.

Further evidence of this phase transition was provided
in~\cite{Koutsoumbas:2006xj}. The QNMs of the electromagnetic
perturbations of the MTZ black hole and its charged generalization
were calculated both analytically and numerically and compared
 with the corresponding QNMs of the vacuum TBH. It was found that there is a change in the slope of the
QNMs as we decrease the value of the horizon radius below a
critical value, and it has been argued that this change signals a
phase transition of a vacuum topological black hole towards the
MTZ black hole with scalar hair.

\section{Stability Analysis}
\label{secst}

To perform the stability analysis of the MTZ black hole it is more convenient to work in the Einstein frame (eqs.~(\ref{eqxI}) and (\ref{eqxV})).
Henceforth, we shall work in units in which the radius of the boundary is $l=1$.

The MTZ black hole metric can be written in the form
\begin{equation}
ds^{2}=-\frac{f}{h^{2}}dt^{2}+\frac{dr^{2}}{f}+r^{2}d\sigma^{2}~,
\end{equation}
where
\bea f &=& f_0(r) = \left[ r^2 - \left( 1 + \frac{r_0}{r} \right)^2 \right] \left( 1 + \frac{r_0^2}{(r + 2r_0)(r+r_0)} \right)^2~, \nonumber\\
h&=& h_0(r) = \left( 1 + \frac{r_0^2}{(r + 2r_0)(r+r_0)} \right)
\frac{r+r_0}{\sqrt{r(r+2r_0)}}~, \eea and the scalar field solution
reads \be \hat\phi = \phi_0 (r) = \pm \sqrt{\frac{3}{4\pi G}}
\tanh^{-1} \frac{r_0}{r+r_0}~. \ee Since the potential (\ref{eqxV})
is an even function, the sign of $\phi_0$ is arbitrary; we shall
choose it so that the leading coefficient in the large $r$ expansion
is positive. Thus, for large $r$, we obtain \be\label{baound}
\hat\phi (r) = \frac{\alpha}{r} + \frac{c\alpha^2}{r^2} + \dots \ee
where \be\label{eqxc} \alpha =\alpha_0 = \sqrt{\frac{3}{4\pi G}}
|r_{0}| \ \ , \ \ \ \ c = - \sqrt{\frac{4\pi G}{3}} \mathrm{sign}
(r_{0})~. \ee Solutions to the Einstein equations with boundary
conditions (\ref{eqxc}) have been found in the case of spherical
horizons and shown to be unstable \cite{Hertog:2004bb}. In that
case, for $\alpha >0$, it was shown that $c<0$ always and the hairy
black hole had positive mass. In our case, we also have $c<0$ when
the black hole has positive mass. However, we obtain $c>0$ when the
black hole has negative mass, which is never the case with spherical
horizons.

We shall now show that in the former case the black hole is unstable
(losing its hair to turn into a TBH) whereas in the latter, it is
stable, as expected from thermodynamic considerations (see section
\ref{sec3}).

To this end,
we apply
the perturbation~\cite{Hertog:2004bb}
\begin{equation}
f(r,t)=f_{0}(r)+f_{1}(r)e^{\mu t} ,\,\,\,\,\, h(r,t)=h_{0}(r)+h_{1}(r)e^{\mu t},\,\,\,\,\,
\phi(r,t)=\phi_{0}(r)+\frac{\phi_{1}(r)}{r} e^{\mu t}~.\,\,\label{perts}
\end{equation}
with $\mu >0$ for an instability to develop.

The field equations give a Schr\"odinger-like wave equation for the scalar perturbation,
\begin{equation}
-\frac{d^{2}\phi_{1}}{dr^{2}_{*}}+\mathcal{V}\phi_{1}=-\mu^{2}\phi_{1}~, \label{scal1}
\end{equation}
where we defined the tortoise coordinate \be
\frac{dr_{*}}{dr}=\frac{h_{0}}{f_{0}} \ee and the effective
potential is given by \be \mathcal{V} = \frac{f_0}{h_0^2} \left[ -
\frac{1}{2} (1+r^2 {\phi_0'}^2) {\phi_0'}^2 f_0 + (1-r^2
{\phi_0'}^2 ) \frac{f_0'}{r} + 2r\phi_0' V'(\phi_0) + V''(\phi_0)
\right]~, \ee
as in the case of a spherical horizon \cite{Hertog:2004bb}.

Regularity
of the scalar field at the horizon of the MTZ black hole ($r\to r_+$) requires  the boundary
conditions \cite{Hertog:2004bb} \be \phi_1 = 0 \ \ , \ \ \ \ (r-r_+) \phi_1' =
\kappa\mu \phi_1\ \ , \ \ \ \ \kappa > 0 ~. \ee For a given $\mu >0$, they uniquely
determine the wavefunction.

\begin{figure}[!t]
\centering
\includegraphics[scale=0.3,angle=-90]{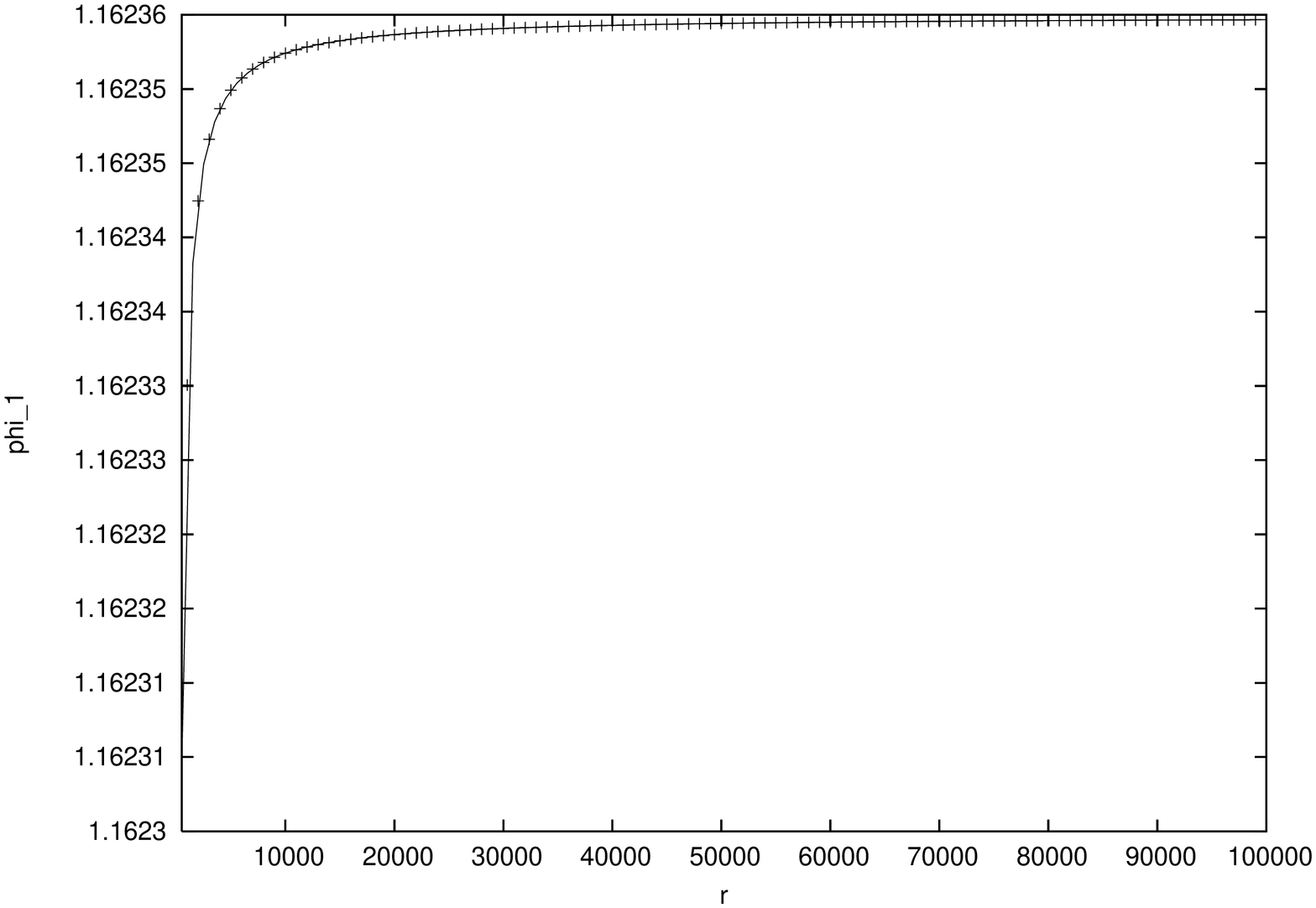}
\includegraphics[scale=0.3,angle=-90]{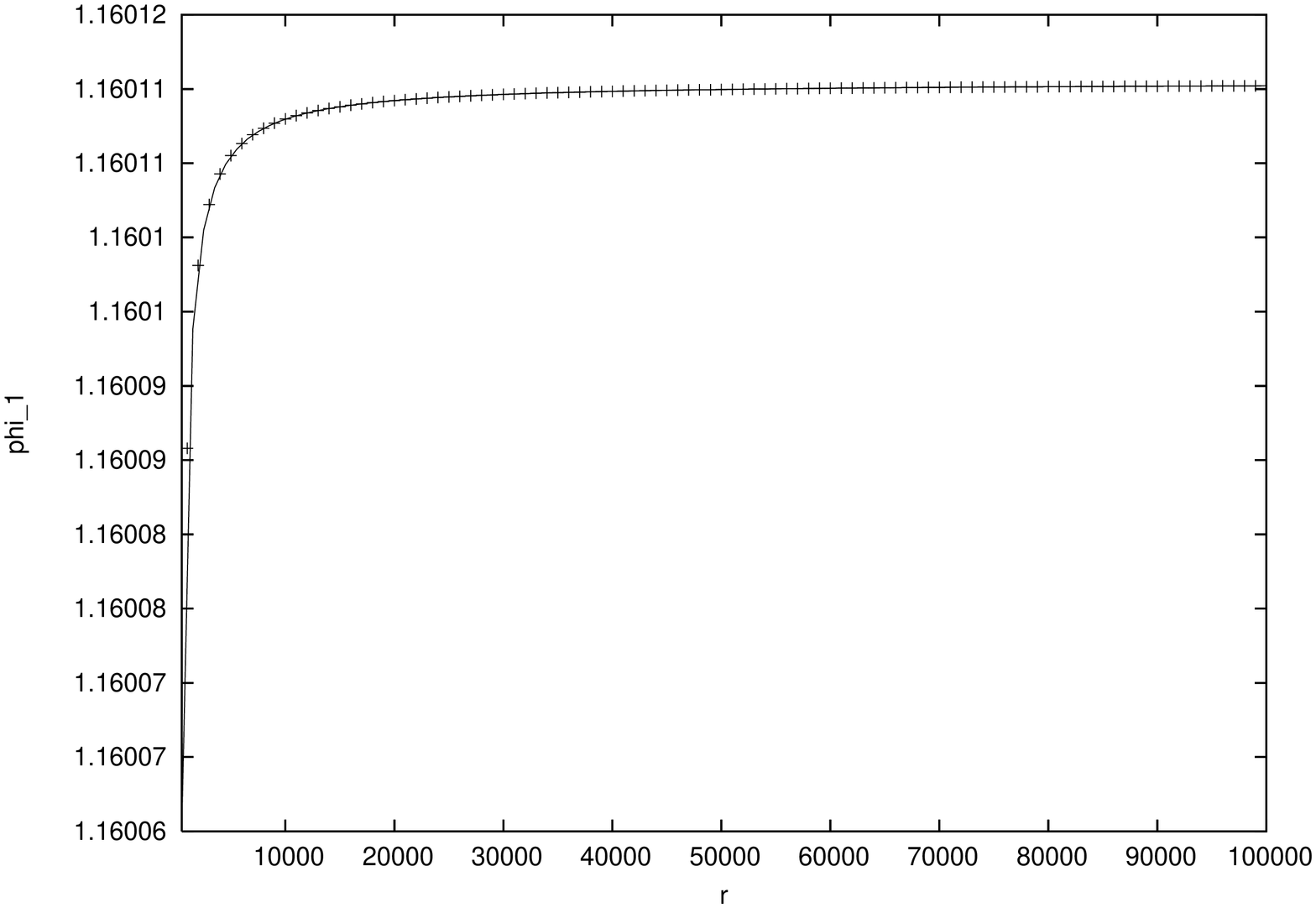}
\caption{The perturbation $\phi_1$ versus $r$ for $\mu=0.02$
and negative ($r_0=-0.01,$ left graph) or positive ($r_0=-0.01,$
right graph) value for $r_0.$} \label{BA}
\end{figure}

At the boundary, the wave equation is approximated by \be -
\frac{d^2\phi_1}{dr_*^2} + 5r_0^2 \phi_1 = -\mu^2 \phi_1~, \ee with
solutions \be \phi_1 = e^{\pm E r_*} \ \ , \ \ \ E = \sqrt{\mu^2 + 5
r_{0}^2}~, \ee where $ r_* =  - \frac{1}{r} + \dots $~. Therefore,
for large $r$, \be \phi_1 = A + \frac{B}{r} + \dots \label{fitAB}\ee
To match the boundary conditions (\ref{baound}) with $c$ given by
(\ref{eqxc}) and $\phi$ expanded as in (\ref{perts}), we need
\be\label{eqx33} \frac{B}{A} =- 2 r_{0}~. \ee Since the wavefunction
has already been determined by the boundary conditions at the
horizon and therefore also the ratio $B/A$, this is a constraint on
$\mu$. If (\ref{eqx33}) has a solution, then the black hole is
unstable. If it does not, then there is no instability of this type
(however, one should be careful  with non-perturbative
instabilities, see \cite{bibxxx}). In figure \ref{BA} we give sample
results for both negative and positive values of $r_0.$ If one fits
these functions as in (\ref{fitAB}), it is obvious that
$\frac{B}{A}$ will be negative in both cases. If $r_0$ is positive,
equation (\ref{eqx33}) may have a solution for some value of $\mu,$
so there is an instability. On the contrary, if $r_0$ is negative,
then $- 2 r_{0}>0$ and equation (\ref{eqx33}) has clearly no
solution, so the black hole is stable.

\section{The Dual Superconductor}
\label{sec4}

Using the results of previous sections we will discuss the dual superconductor on the boundary of the exact
supergravity solution.
Recall that for $T<T_0$, a condensate
forms ($\phi\ne 0$), the field equations have as a solution the MTZ black
hole (\ref{MTZconf}) and the scalar field is given by (\ref{Psiconf}).
For $T>T_0$, $\phi= 0$, no
condensate forms and the field equations have as a solution
the TBH of (\ref{linel}).

The metric on the boundary is given by (\ref{eqnew}) and its curvature is $-1/l$ (hyperbolic manifold).
As in the previous section, we shall keep working in units in which the radius of the boundary is $l=1$.

It should be noted that the mechanism of condensation of the scalar
field here is different than the mechanism of condensation of the
dual superconductor in the case of a black hole of flat
horizon~\cite{Hartnoll:2008vx}. In the latter case the scalar field
condenses because an Abelian-Higgs mechanism was in operation. In
our case, the condensation of the scalar field has a geometrical
origin and is due entirely to its coupling to gravity.

The energy of the superconductor which is dual to the hairy black hole solution is (eq.~(\ref{relations1}))
\be E_s(T) = M_{MTZ} =
-\frac{\pi\sigma }{4 G} \left(  T_0^2 - T^2 \right) ~, \ee which is
negative for temperatures below the critical point $T_0$. However,
notice that the negative contribution is entirely due to the
Casimir energy. If we subtract the energy at $T=0$, we obtain \be
E_s(T) - E_s(0) = \frac{\pi\sigma }{4 G}\ T^2, \ee which is a
positive quantity at all temperatures.

Moreover, the heat capacities in the normal and superconducting
phases, dual to the TBH and MTZ solutions, respectively, are \be C_n
= T \frac{\partial S_{TBH}}{\partial T} \ \ , \ \ \ \ C_s = T
\frac{\partial S_{MTZ}}{\partial T}~. \ee Using the explicit form of
the entropy (eqs.~(\ref{relations2}) and (\ref{relations1})), near
zero temperature we have \be\label{eqheatc} C_n \approx
\frac{\pi\sigma }{3\sqrt 3 G}\ T \ \ , \ \ \ \ C_s \approx
\frac{\pi\sigma }{2G}\ T \ , \ee therefore both heat capacities
vanish linearly with temperature as $T\to 0$. Such a power law
dependence was also observed in the case of charged holographic
superconductors in flat space \cite{Hartnoll:2008kx}, although the
numerical value of the power could not be determined in that case.
This departure from the typical behaviour of $s$-wave
superconductors  (exponential suppression of the heat capacity as
$T\to 0$) may be due to the presence of a Goldstone mode
\cite{Hartnoll:2008kx}.

To the exact gravity backgrounds we shall now apply an
electromagnetic perturbation. In the case without a condensate (TBH) the
wave equation for perturbing the vector potential
reads~\cite{Koutsoumbas:2006xj} \be
f_{TBH}(f_{TBH}A')'+\left(\omega^{2}-\frac{\xi^{2}+1/4}{r^{2}}f_{TBH}\right)A=0~,\label{weqtbh}\ee
where $A$ is an appropriately defined component of the vector potential.

The wavefunction of the lowest harmonic corresponds to the lowest eigenvalue in
the compact hyperbolic space $\Sigma$ (eq.~(\ref{eqSigma})) given by \be
\xi^{2}+\frac{1}{4}=0~.\label{eigenvalue}\ee Defining the tortoise
coordinate
$$r_{*}=-\int^{\infty}_{r}\frac{dr'}{f_{TBH}(r')}~,$$ eq.~(\ref{weqtbh}) is solved to give (with arbitrary normalization)
\be\label{eqBC} A(r)=e^{-i\omega r_{*}} \ee which obeys the
correct conditions at the horizon. For large $r$ we have
$r_{*}=-1/r+\mathcal{O}(r^{-3})$ therefore the vector potential at
large radius behaves asymptotically as \be A = A^{(0)} +
\frac{A^{(1)}}{r} + \dots \label{expan}\ee According to the
AdS/CFT correspondence, the dual source and the expectation value
for the current in the same direction are given respectively by
\be A = A^{(0)} \,, \qquad \langle J \rangle = A^{(1)} \,. \ee
Then the conductivity is \be\label{eq:conductivity} \sigma(\omega)
= \frac{\langle J \rangle}{E} = - \frac{ \langle J \rangle}{\dot
A} = -\frac{ \langle J \rangle}{i\omega A} =
\frac{A^{(1)}}{i\omega A^{(0)}} \,. \ee We have $A^{(0)}=1$,
$A^{(1)}=i\omega$ and therefore the conductivity
(\ref{eq:conductivity}) reads \be
\sigma(\omega)=\frac{A^{(1)}}{i\omega
A^{(0)}}=1~.\label{conduct}\ee According to the discussion in the
previous section, if the temperature is above the critical
temperature $T_{0}= \frac{1}{2\pi }$ (eq.~(\ref{eqTcr}) setting the radius $l=1$), the most probable configuration is the
vacuum TBH. Then relation~(\ref{conduct}) tell us that  the
boundary conducting theory is in the normal phase, as expected.

If the temperature is below the critical temperature, the
vacuum TBH acquires hair and a condensate forms.
 In this
case, (\ref{Psiconf1}) can be expanded as \be
\Psi=\frac{\Psi^{(1)}}{r}+ \frac{\Psi^{(2)}}{r^{2}}+\dots ~, \ee where
\be \Psi^{(1)}=-\sqrt{\frac{3}{4 \pi G}}\, r_0\,,\,\,\,\,
\Psi^{(2)}=\sqrt{\frac{3}{4 \pi G}}\, r_0^{2}~.\ee
The two non-vanishing leading coefficients lead to condensates of two dual scalar operators $\mathcal{O}_i$ ($i=1,2$), respectively,
\be\label{eqO1} \langle \mathcal{O}_i \rangle = \sqrt 2 \Psi^{(i)} ~,\quad i=1,2~. \ee
Unlike in the case of a flat horizon~\cite{Hartnoll:2008vx}, the existence of two condensates does not imply an instability, as was shown in section \ref{secst}.

\begin{figure}[!t]
\centering
\includegraphics[scale=0.8,angle=0]{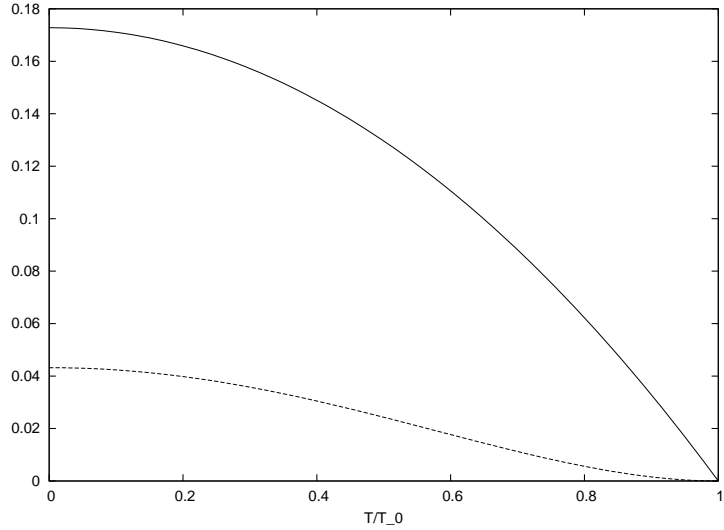}
\caption{The condensates $\sqrt G \langle\mathcal{O}_1\rangle$ (upper curve) and $\sqrt G \langle\mathcal{O}_2\rangle$ (lower curve) as functions of $T/T_0$ (eq.~(\ref{eqO})).}
\label{figO}
\end{figure}

The two condensates (\ref{eqO1}) are, respectively,
\be\label{eqO} \langle
\mathcal{O}_{1}\rangle=\sqrt{\frac{3\pi^3}{2 G}}\, (T_0^2 -T^2)
\,,\,\,\,\,\,\langle \mathcal{O}_{2}\rangle=\sqrt{\frac{3\pi^7}{2
 G}}\,(T_0^2 -T^2)^{2}~, \ee where we used
(\ref{relations1}) and (\ref{horizon}) to express the condensates as
functions of temperature.  Near the critical temperature $T_{0}$ the
condensates behave respectively as \be \langle
\mathcal{O}_{1}\rangle \simeq \sqrt{\frac{3}{8\pi G}}\,\left( 1-
\frac{T}{T_{0}} \right) \,,\,\,\,\,\,\langle
\mathcal{O}_{2}\rangle\simeq \sqrt{\frac{3}{32 \pi G}}\,\left( 1-
\frac{T}{T_{0}} \right)^{2}~. \ee The two condensates are plotted as
functions of temperature in figure~\ref{figO}. Observe that the
condensates $\langle\mathcal{O}_{1}\rangle,\,
\langle\mathcal{O}_{2}\rangle$  have a different temperature
dependence than the corresponding operators in the dual
superconductor of a black hole with flat
horizon~\cite{Hartnoll:2008vx}. This behaviour of the condensates is
reminiscent of materials containing impurities \cite{Skalski}.

Applying the electromagnetic perturbations to this background, the
wave equation for the vector potential reads \be
f_{MTZ}\left(f_{MTZ}A'\right)'+\left(\omega^{2}-\frac{\xi^{2}
+1/4}{r^{2}}f_{MTZ}-q^{2}\Psi^{2}f_{MTZ}\right)A=0~.\label{weqmtz}\ee
If we use the lowest angular eigenvalue (\ref{eigenvalue}), this
simplifies to \be
f_{MTZ}\left(f_{MTZ}A'\right)'+\left(\omega^{2}-q^{2}\Psi^{2}f_{MTZ}\right)A=0~.\label{simpleweqmtz}\ee
Eq.~(\ref{simpleweqmtz}) can not be solved analytically in general. However, we
can solve this equation for weak coupling $q^{2}$ using
perturbation theory. Also, a numerical analysis of
(\ref{simpleweqmtz}) will be presented in the next section. For
$q=0$ the solutions are $e^{\pm i\omega r_{*}}$ where
$$r_{*}=-\int^{\infty}_{r}\frac{dr'}{f_{MTZ}(r')}~.$$ The
acceptable solution is $e^{- i\omega r_{*}}$. Using first-order
perturbation theory, we obtain the solution \be A=e^{- i\omega
r_{*}}+\frac{q^{2}}{2i\omega}e^{ i\omega
r_{*}}\int^{r}_{r_{+}}dr'\Psi^{2}(r')e^{-2 i\omega r_{*}}
-\frac{q^{2}}{2i\omega}e^{- i\omega
r_{*}}\int^{r}_{r_{+}}dr'\Psi^{2}(r')~.\ee The lowest limit of the
integration has been chosen at $r_{+}$ to ensure correct behaviour
at the horizon. Expanding $A$ in the large $r$ limit (equation
(\ref{expan})), we get \be A^{(0)}=\left. A \right|_{r\rightarrow
\infty}=1+\frac{q^{2}}{2i\omega}\int^{\infty}_{r_{+}}dr\Psi^{2}(r)\left[
e^{-2 i\omega r_{*}}-1\right]~, \ee \be
A^{(1)}= \left. -r^{2}\frac{dA}{dr} \right|_{r\rightarrow
\infty}=i\omega-\frac{q^{2}}{2}\int^{\infty}_{r_{+}}dr\Psi^{2}(r)\left[
e^{-2 i\omega r_{*}}+1\right]~.\ee Then, the conductivity to
first-order in $q^{2}$ is \be
\sigma(\omega)=\frac{A^{(1)}}{i\omega
A^{(0)}}=1-\frac{q^{2}}{i\omega}\int^{\infty}_{r_{+}}dr\Psi^{2}(r)
e^{-2 i\omega r_{*}}~.\label{conductivitymtz}\ee
The superfluid
density is the coefficient of the delta function of the real part
of the conductivity \be \Re[\sigma(\omega)] \sim \pi n_s
\delta(\omega) \,, \ee which is also the coefficient of the pole
in the imaginary part
\be\label{eqimomega} \Im[\sigma(\omega)] \sim \frac{n_s}{\omega} ~,\quad
\omega \to 0~. \ee
Therefore using
(\ref{conductivitymtz}) we obtain for the superfluid density \be
n_{s}=q^{2}\int^{\infty}_{r_{+}}dr\Psi^{2}(r)=\frac{3q^{2}}{4\pi
G}\, \frac{r_0^{2} }{r_{+}+r_0 }~, \ee or in terms of the temperature, \be\label{eqalpha} n_{s} (T) = \alpha\left(T_{0}-T\right)^{2} ~,\quad \alpha = \frac{3\pi
q^{2}}{4G}~. \ee
Notice that near $T=0$,
\be\label{eqnsa} n_s(0) - n_s(T) \approx \frac{\alpha}{\pi} T \ee
matching the low temperature behaviour of the heat capacity (\ref{eqheatc}).
Such power law behaviour was also observed in charged holographic superconductors in flat space \cite{Hartnoll:2008kx}. Once again, we see an indication that a Goldstone mode is in play.

The normal, non-superconducting,
component of the DC conductivity is given by
\be n_n = \lim_{\omega \to 0} \Re
[\sigma(\omega)] \,. \label{normalConductivity}\ee
We obtain
\be
\ln n_{n}=2q^{2}\int^{\infty}_{r_{+}}dr\Psi^{2}(r)r_{*}~. \ee
Integrating by parts, we obtain \be \ln n_{n} = \frac{3 q^{2}}{2\pi
G} r_0^{2}h(r_{+})~,\label{normalcond1}\ee where \be
h(r_{+})=-\frac{1}{r_0^{2}}\ln(r_{+}+r_0)-\frac{1}{(r_{+}+r_0)}\,\sum^{3}_{i=1}
\frac{(r_{+}-r_{i})\ln(r_{+}-r_{i})}{f'_{MTZ}(r_{i})(r_{i}+r_0)}\label{hfunction}\ee
and $r_{i}$ are the three horizons of MTZ black hole given in
(\ref{horizon}). At low temperatures we have $r_0\rightarrow
-\frac{1}{4}$ and the function $h(r_{+})$ (\ref{hfunction}) is
approximated by \be   h(r_{+})\simeq -\frac{1}{(r_{+}+r_0)}\,
\frac{(r_{+}-r_{-})\ln(r_{+}-r_{-})}{f'_{MTZ}(r_{-})(r_{-}+r_0)}\simeq
8 \ln T~.\ee Then the normal component of the DC conductivity
(\ref{normalcond1}) becomes \be \ln n_{n}\simeq \frac{3q^{2}}{4 \pi
G}\ln T~.\label{loggap} \ee
leading to the low temperature behaviour \be n_{n}\sim
T^{\gamma}\,, \,\,\,\,\, \gamma=\frac{3q^{2}}{4\pi
G}~.\label{polygap} \ee The approximation of the  normal component
of the DC conductivity (\ref{polygap}) has a milder behaviour than
the dual superconductor in the case of a black hole of flat
horizon~\cite{Hartnoll:2008vx} in which $n_{n}$ exhibits a clear
gap behaviour. The  behaviour we observe here of the boundary
conducting theory can be found in materials with paramagnetic
impurities \cite{Skalski} and in unconventional superconductors
like the chiral p-wave superconductor \cite{Furusaki}.

These analytical results are supported by a numerical investigation of equation (\ref{simpleweqmtz}) which  will be discussed in the next section.

\section{Numerical Results}
\label{sec5}

In this section we discuss the numerical solution of the wave equation (\ref{simpleweqmtz}) in the interval
$r\in [r_+, \infty)$ and compare it with the analytical results obtained above using perturbation theory.
We shall be working in units in which the radius of the hyperbolic space is $l=1$, so that, e.g., distances will be given in units of $l$ and frequencies and temperatures will be in units of $1/l$.

The boundary condition at the horizon ($r=r_+$) is given by eq.~(\ref{eqBC})
which implies
\be\label{eqBC1} A^\pr(r) = -\fr{i \omega}{f_{MTZ}(r)}A(r)~.\ee
Since, by definition, $f_{MTZ}(r_+)=0,$ we applied the boundary
condition (\ref{eqBC1}) at $r=r_++\epsilon,$ where $\epsilon$ is a
small positive quantity. In the following calculations the value
$\epsilon=\fr{r_+}{10000}$ has been used. We integrated up to
$r=R\equiv 5000 r_+$. We solved the wave equation for $r_+ \le 1$
($T\le T_0$) since we know that in this temperature range the
topological black hole acquires hair, while  for $r_+ > 1$ the
vacuum TBH is energetically favorable.

By curve fitting the solution of the wave equation, we calculated
the coefficients $A^{(0)}$ and $A^{(1)}$ defined through
(\ref{expan}). We then deduced the conductivity $\sigma$ using
(\ref{eq:conductivity}). In the limit $\omega\to 0$, the
conductivity yields the densities of the superfluid and normal
components via (\ref{eqimomega}) and (\ref{normalConductivity}),
respectively. We chose $\omega=0.001$ for our numerical
calculations.


\begin{figure}[!t]
\centering
\includegraphics[scale=0.3,angle=-90]{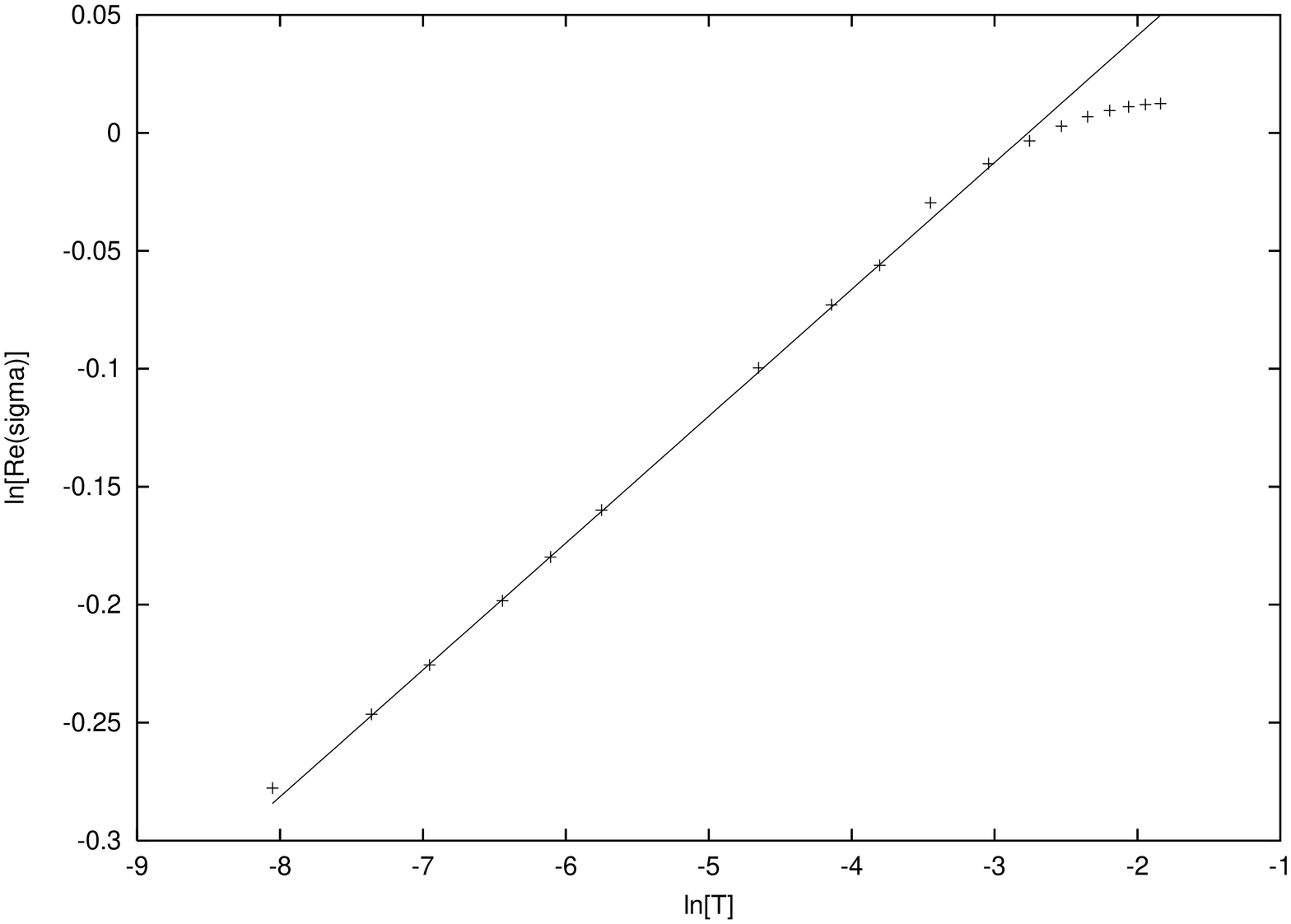}
\includegraphics[scale=0.3,angle=-90]{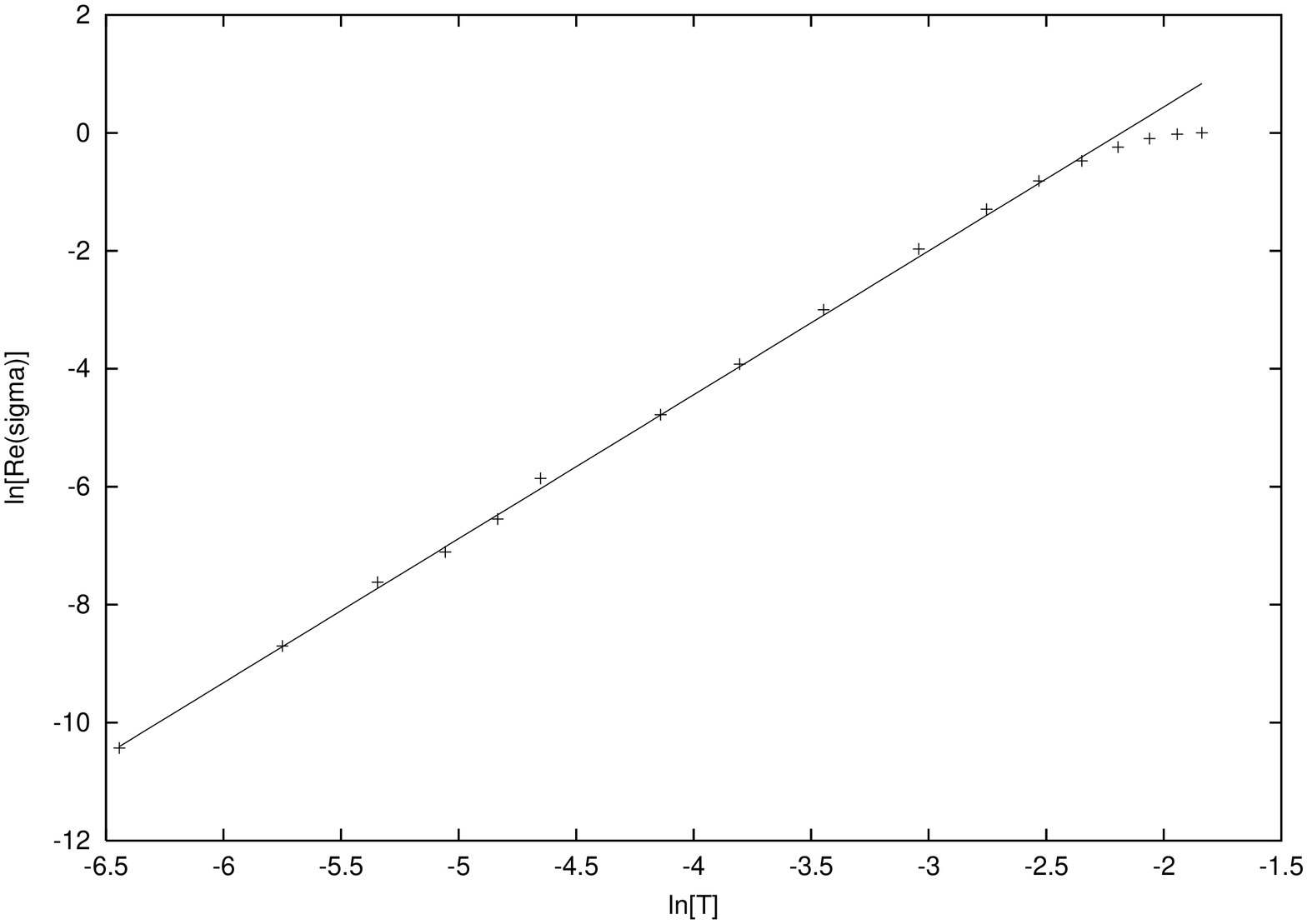}
\caption{The logarithm of the normal fluid density as a function
of the logarithm of the temperature for $q/\sqrt G=0.5$ (left) and
$q/\sqrt G=5.0$ (right). The solid lines represent the fits
$\ln n_n =0.0538 \ln T +0.149$ (left) and $\ln n_n=2.45 \ln T
+5.3$ (right).} \label{q=0.5_nn}
\end{figure}

\begin{figure}[!b]
\centering
\includegraphics[scale=0.5,angle=-90]{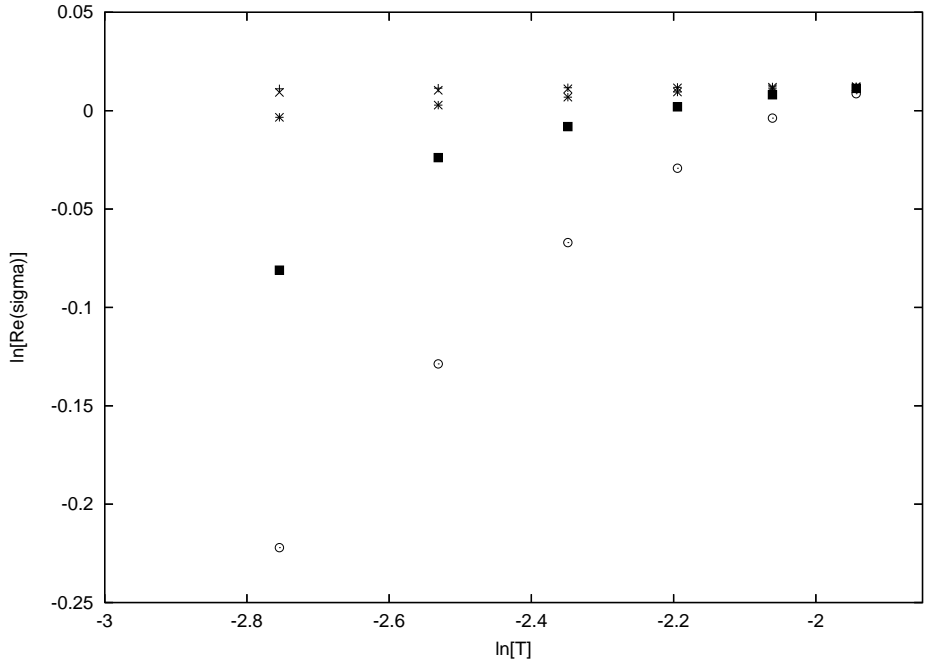}
\caption{The logarithm of the real part of the conductivity as a
function of the logarithm of the temperature for $q/\sqrt G=0.1$ (crosses,
the uppermost symbols),$\, 0.2,0.5,1.0,2.0.$ } \label{comp_nn}
\end{figure}

On the basis of the analytic results in section \ref{sec4} we
expect that at low temperature, the normal fluid density can be
expanded as \be\label{eqfitn} \ln n_n = \gamma \ln T + \delta +
\dots \ee whereas near the critical temperature, the superfluid
density is expanded as
\be\label{eqfits} n_s = \alpha (T-T_0)^2+ \dots \ee We therefore
fit the data accordingly. The results for $\ln n_n$ at $q/\sqrt
G=0.5$ and $q/\sqrt G=5.0$ are shown in figure \ref{q=0.5_nn}. The
fit (\ref{eqfitn}) has taken into account only the points below
$\ln T = -5$. This is why the data points at higher temperatures
($\ln T > -5$) lie below the fit.

Figure \ref{comp_nn} contains data for various values of the
charge $q$. We observe that the asymptotic slope of the curve as $T\to 0$
increases with the charge.
Table \ref{table1} contains numerical values of the slope
obtained through the fit (\ref{eqfitn}) and compares them
with their analytical counterparts (\ref{polygap}).

\begin{figure}[!t]
\centering
\includegraphics[scale=0.3,angle=-90]{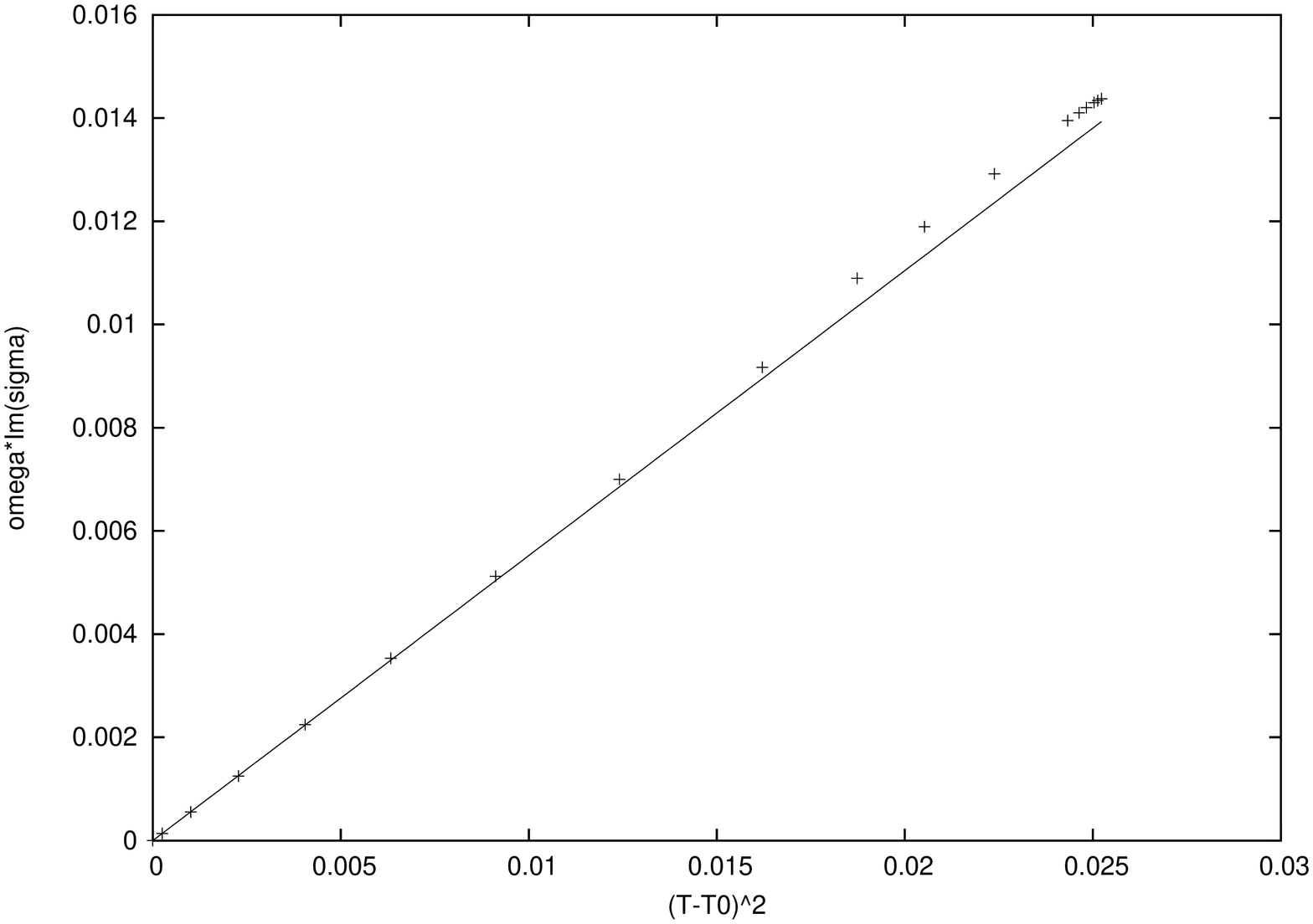}
\includegraphics[scale=0.3,angle=-90]{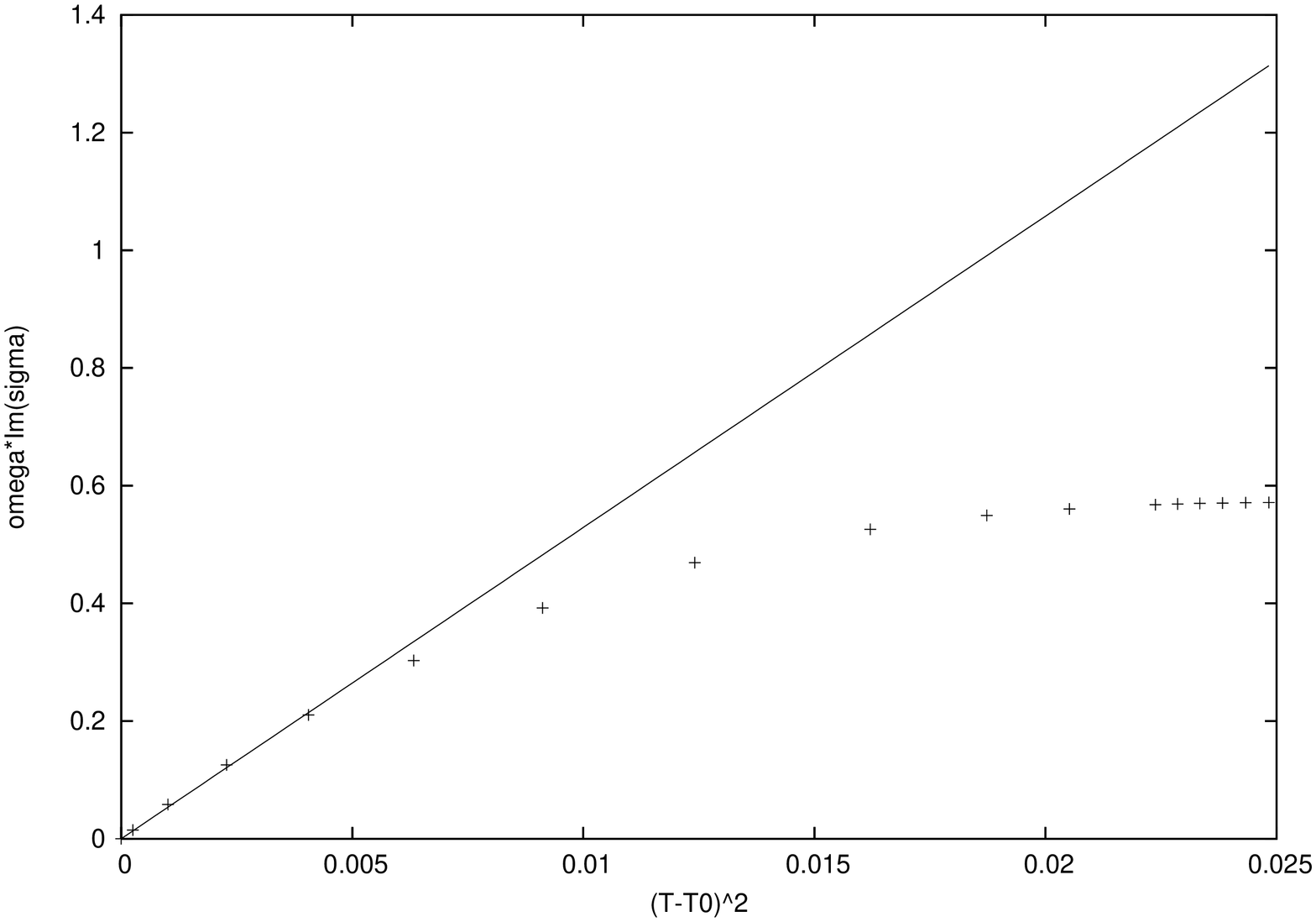}
\caption{Superfluid density as a function of $(T-T_0)^2$ for
$q/\sqrt G=0.5$ (left) and $q/\sqrt G=5.0$ (right). The solid lines
represent the fits $n_s=0.552 (T-T_0)^2$ (left) and
$n_s=52.9 (T-T_0)^2$ (right).} \label{q=0.5_ns}
\end{figure}

We present our results for the superfluid density $n_s$ {\it vs}
temperature in figure \ref{q=0.5_ns} for $q/\sqrt G =0.5, 5.0$.
The fit has been done in the region of temperatures
$(T-T_0)^2<0.005,$ so it does not represent accurately the data at
temperatures away from this region. Again, table \ref{table1}
contains numerical values of the slope obtained through the fit
(\ref{eqfits}) and compares them with their analytical
counterparts (\ref{eqalpha}). The agreement is better for $n_s$
than for $n_n.$

\begin{table}
\begin{center}

\begin{tabular}{||c|c|c|c|c||}

\hline

$q/\sqrt G$ & $\gamma_{\mathrm{numerical}}$ & $\gamma_{\mathrm{analytical}}$&
$\alpha_{\mathrm{numerical}}$ & $\alpha_{\mathrm{analytical}}$ \\

\hline

0.1 & 0.0020 &  0.0024 & 0.0225 & 0.024 \\

0.5 & 0.0538 &  0.0597 & 0.552 & 0.589 \\

1.0 & 0.187 &  0.239 & 2.196 & 2.356 \\

2.0 & 0.684 &  0.955 & 8.678 & 9.425 \\

3.0 & 1.325 &  2.15 & 20.35 & 21.21 \\

5.0 & 2.522 &  5.97 & 52.90 & 58.90 \\

\hline

\end{tabular}

\end{center}
\caption{Numerical {\it vs}~analytical results for the normal and
superfluid densities for various values of the charge. The numerical
parameters are obtained through the fits (\ref{eqfitn}) and
(\ref{eqfits}), respectively.  Their analytical counterparts are
given by (\ref{polygap}) and (\ref{eqalpha}),
respectively.}\label{table1}
\end{table}

At low temperatures, the superfluid density $n_s$ approaches a
constant. After subtracting this constant, we obtain the behaviour
\be\label{eqnsT} n_s(0) - n_s(T) \sim T^\delta \ee where $\delta
\approx 1$ for small charges $q$, in agreement with our earlier
analytic result (\ref{eqnsa}). Sample values of the exponent
$\delta$ are shown in table \ref{table2}. $\delta$ increases as $q$
increases. Similar temperature dependence of $n_s$ was observed in
charged holographic superconductors in flat  space
\cite{Hartnoll:2008kx} indicating the presence of a Goldstone mode.

\begin{table}
\begin{center}

\begin{tabular}{||c||c|c|c||}

\hline

$q/\sqrt G$ & $1$ & $3$&
$5$ \\

\hline

$\delta$ & $1.025\pm 0.007$ &  $1.52\pm 0.03$ & $1.78\pm 0.03$ \\

\hline

\end{tabular}

\end{center}
\caption{The exponent $\delta$ characterizing the low temperature dependence of the superfluid density $n_s$ (eq.~(\ref{eqnsT}))
for various values of the charge.
}\label{table2}
\end{table}

\begin{figure}[!b]
\centering
\includegraphics[scale=0.3,angle=-90]{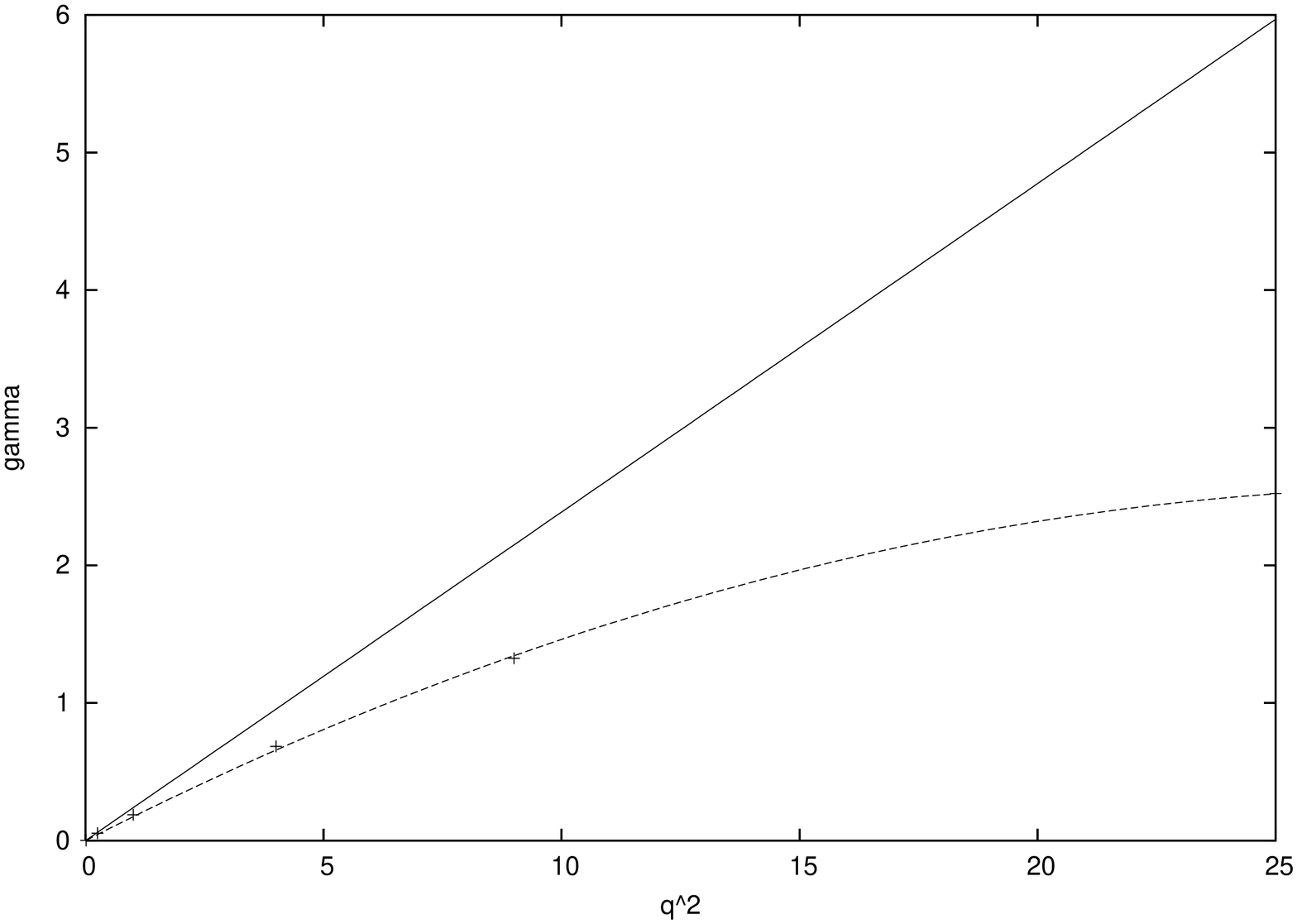}
\includegraphics[scale=0.3,angle=-90]{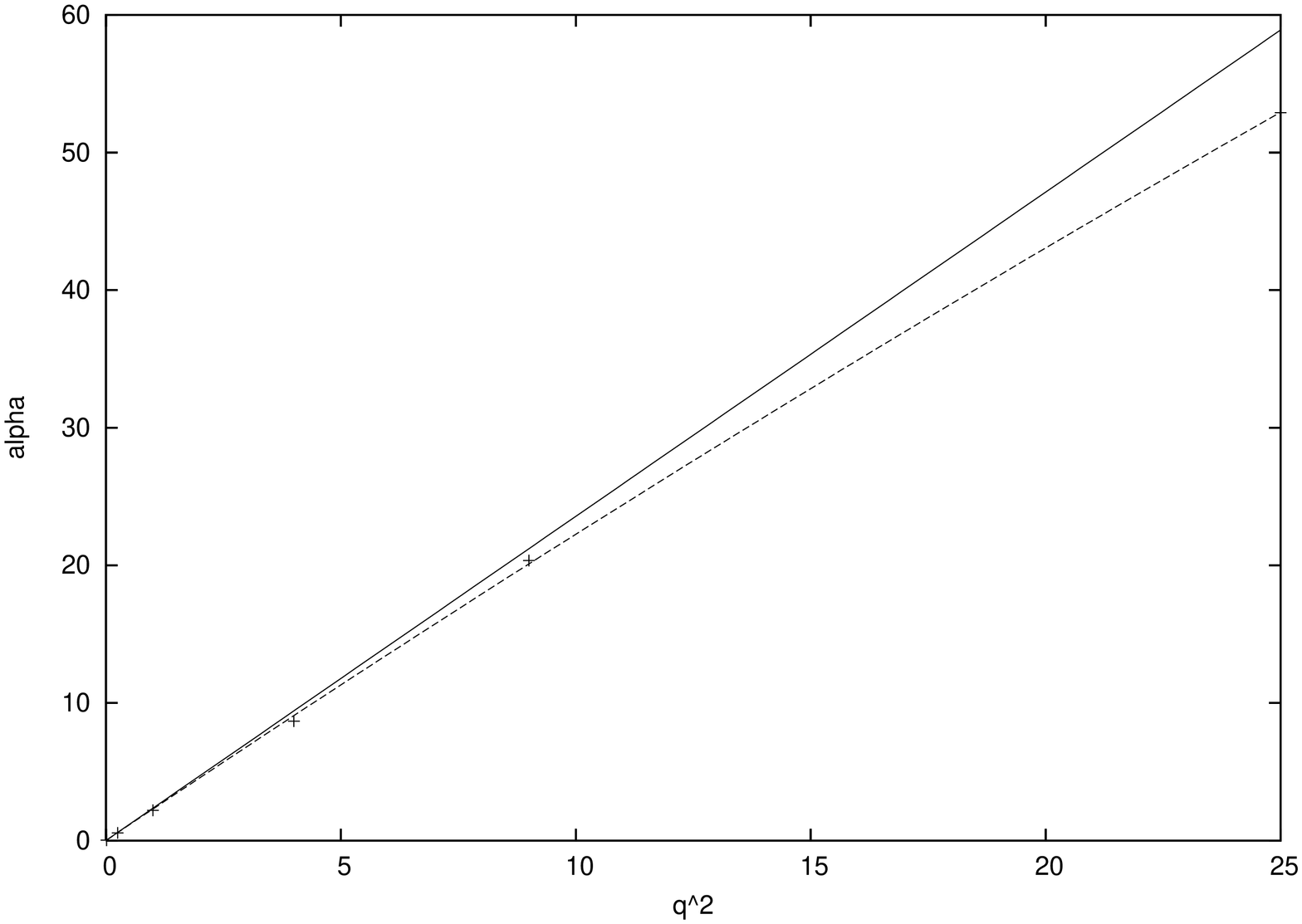}
\caption{Numerical and analytical results for the normal (left) and
superfluid (right) densities {\it vs}~$q^2$.  Numerical data are
fitted by $0.176 q^2 - 0.0030 q^4$ (left) and $2.29 q^2 - 0.007 q^4$
(right).} \label{slopes}
\end{figure}

The results of table \ref{table1} are depicted in figure \ref{slopes}. It is
clear that the agreement between numerical and analytical results
is quite satisfactory for the superfluid density, while serious
discrepancies arise for the normal density as $q$ increases.

\begin{figure}[!b]
\centering
\includegraphics[scale=0.3,angle=-90]{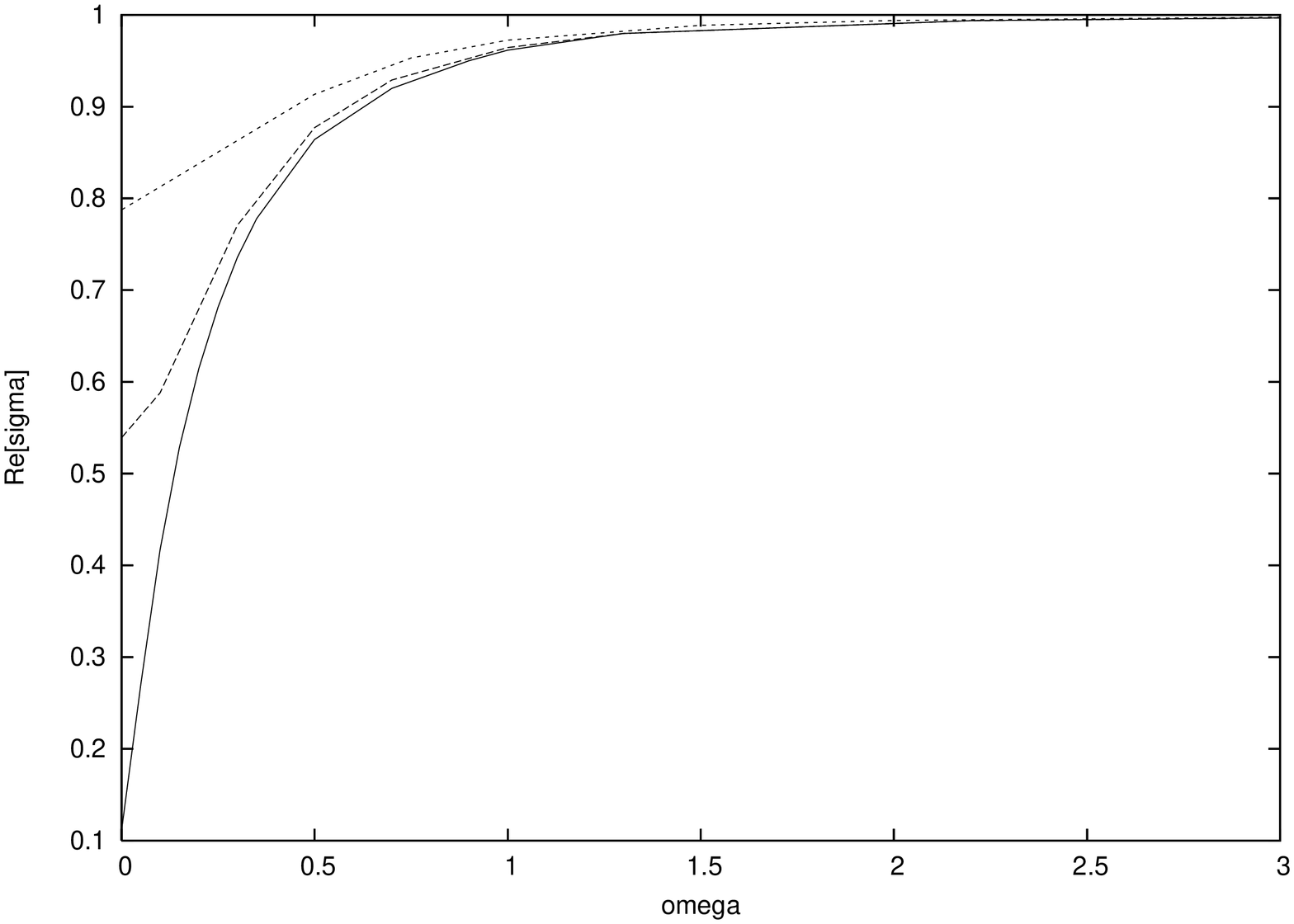}
\includegraphics[scale=0.3,angle=-90]{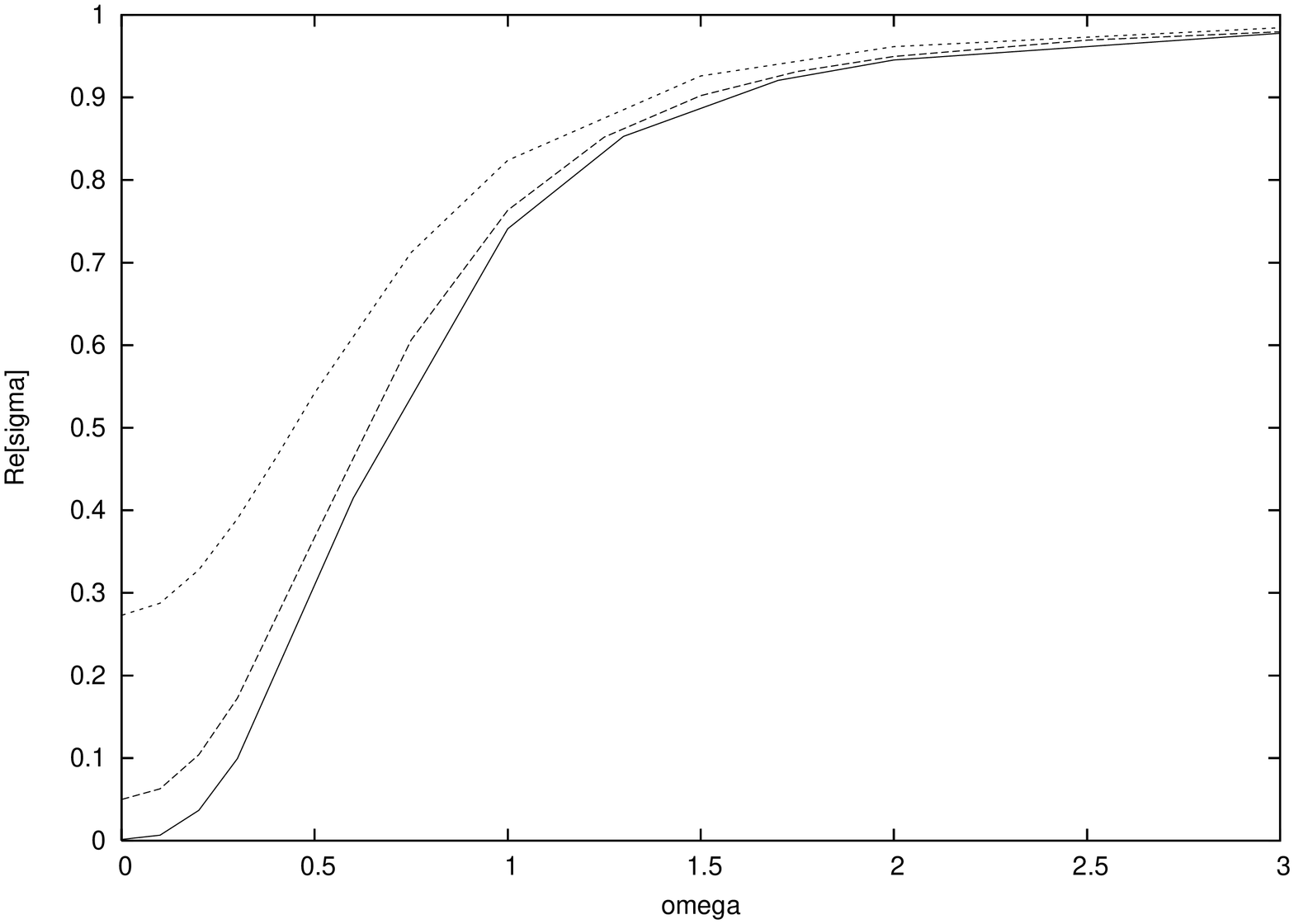}
\caption{The real part of the conductivity {\em vs} $\omega$
for $q/\sqrt G=2$ (left) and $q/\sqrt G=5$ (right) and $T=0.0032, 0.032, 0.064.$ The lowest curve
corresponds to the lowest temperature.} \label{re_q=2}
\end{figure}

We further analyzed the $\omega$ dependence of the transport coefficients.
Figure \ref{re_q=2} contains the real part of the conductivity
{\it vs} $\omega$ for $q/\sqrt G=2,5$ and various values of the temperature.
The lowest value of the temperature yields rather small values for
this real part, while for larger temperatures the real part tends
to the value 1, which is the outcome for the topological black
hole.
Comparing, one may see that the real part of the conductivity becomes
smaller as we increase the charge $q$.
Unfortunately, numerical instabilities also increase and we have not been able to produce reliable numerical results above $q/\sqrt G =5$.
In the cases we studied, it appears that the superconductor is {\em gapless}.
However, a gap is likely to develop above a certain value of the charge $q$, as
indicated by the trend in the graphs as $q$ increases.

\begin{figure}[!t]
\centering
\includegraphics[scale=0.3,angle=-90]{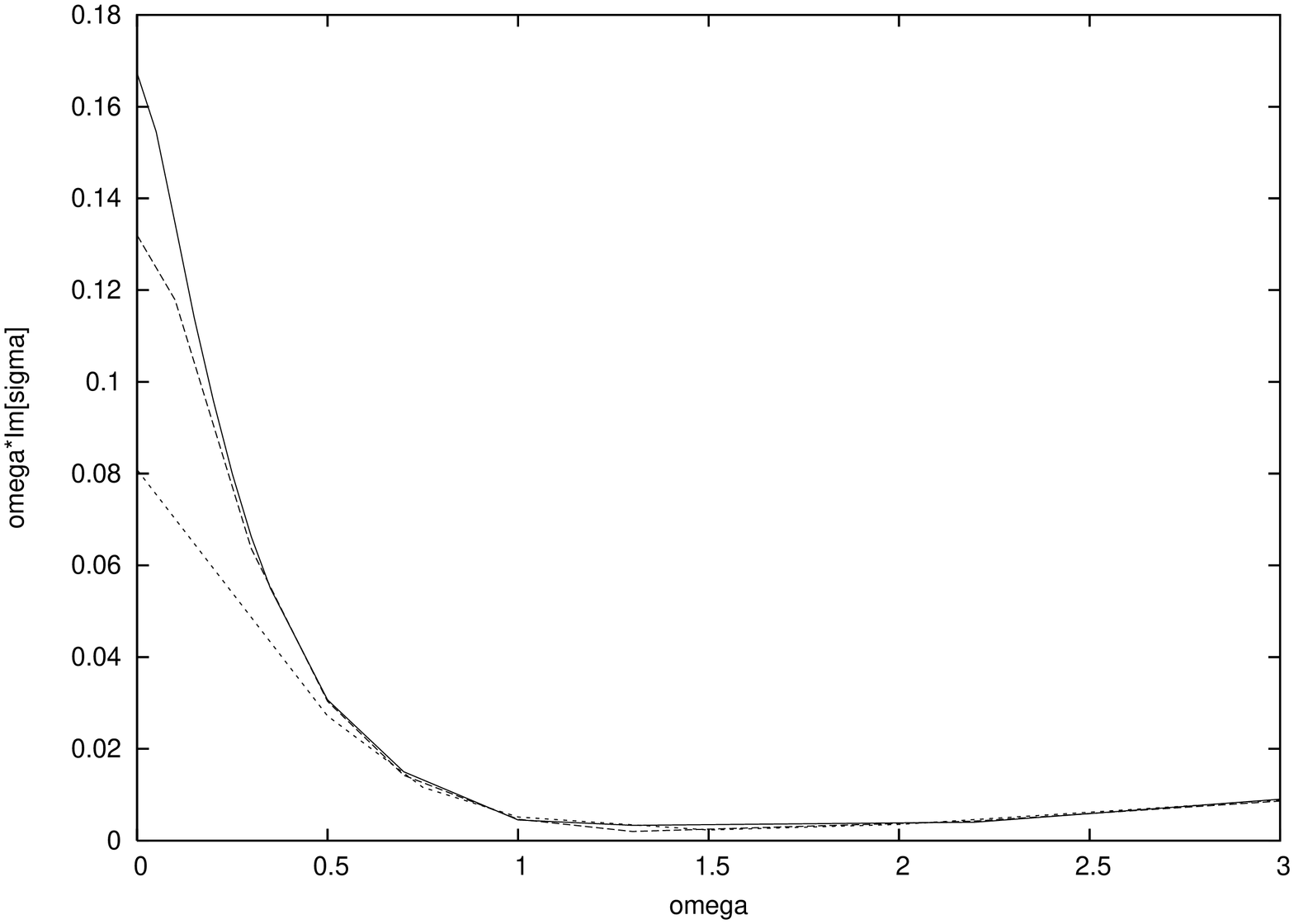}
\includegraphics[scale=0.3,angle=-90]{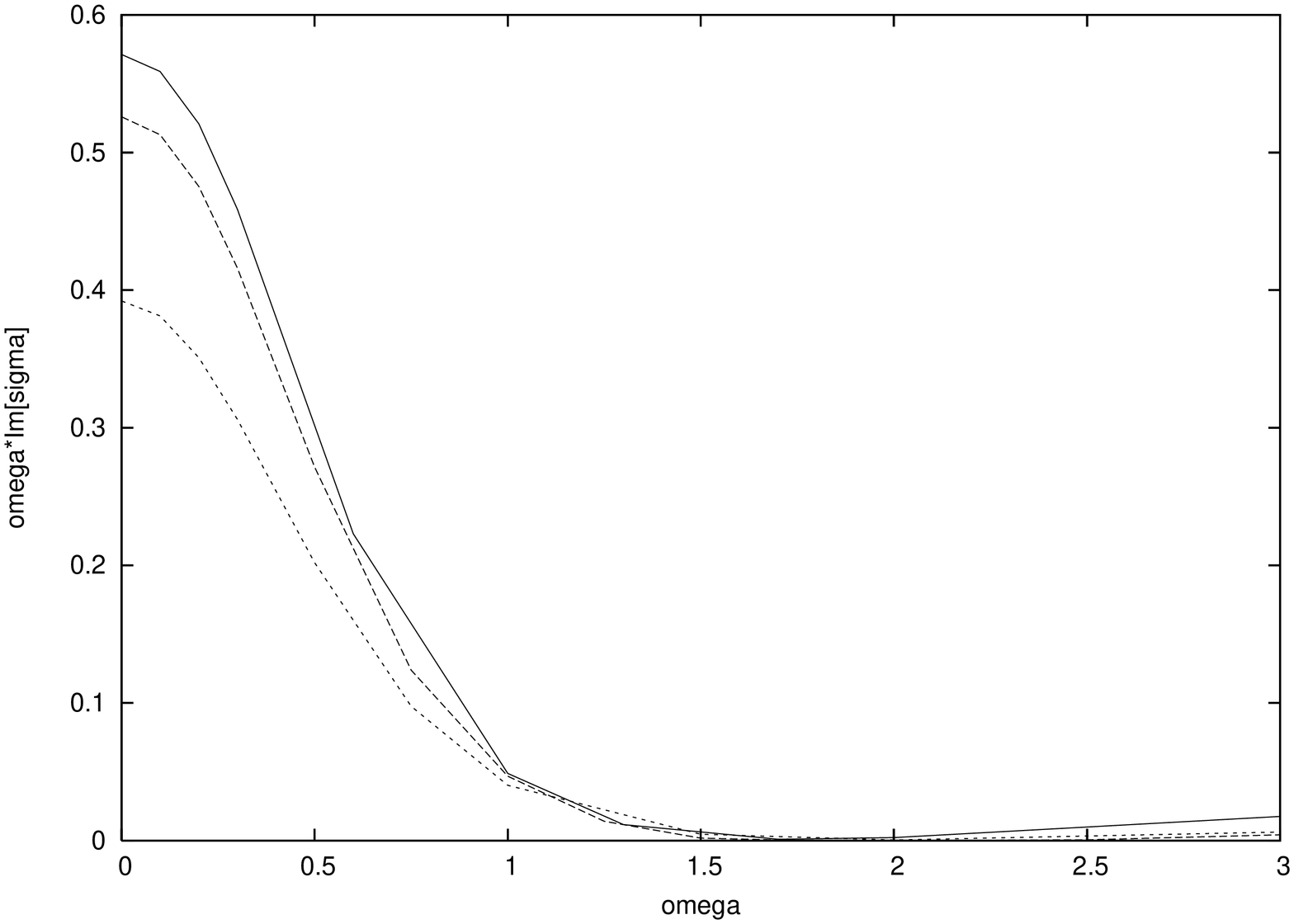}
\caption{The imaginary part of the conductivity multiplied by
$\omega$ {\em vs} $\omega$ for $q/\sqrt G=2$ (left) and $q/\sqrt G=5$ (right) and $T=0.0032, 0.032,
0.064.$ The uppermost curve corresponds to the lowest
temperature.} \label{im_q=2}
\end{figure}

Finally, figure \ref{im_q=2} contains the imaginary part of the
conductivity {\it vs} $\omega$ for $q/\sqrt G =2,5$ and various values of the
temperature.
Some features are more clearly visible in the latter case. The imaginary
part is multiplied by $\omega$ to tame the pole at
$\omega=0.$ The imaginary part seems to vanish at some
frequency, which moves to the left as the temperature increases.

\section{Conclusions}
\label{sec6}

We  presented a model of an exact gravity dual of a gapless superconductor in
which a condensate forms as a result of the coupling of a
charged scalar field to gravity. The charged scalar field
 responsible for the condensation is a solution of the field
 equations~\cite{Martinez:2004nb,Martinez:2005di} and below a critical temperature dresses up a vacuum
 black hole of a constant negative curvature horizon (TBH) with scalar hair.
 Perturbing the background Maxwell field and using the AdS/CFT correspondence,
we determined the conductivity of the boundary theory and analysed
the behaviour of the normal  and superconducting fluid densities
using both analytical and numerical techniques.

The condensation of the scalar field we considered had a purely
geometrical origin being due entirely to its coupling to gravity.
Hairy charged black hole solutions are also known to exist in the
case of a {\it real} scalar field~\cite{Martinez:2005di}. It would
be interesting to extend these solutions to the case of a {\it
complex} (charged) scalar field $\phi$ and analyse the entire range
of parameters labeling the solutions, including the charges of the
black hole and the scalar field. This should yield an interesting
landscape consisting of chargeless as well as charged
superconductors,  as in the flat
case~\cite{Hartnoll:2008vx,Hartnoll:2008kx}. Work in this direction
is in progress.


\section*{Acknowledgments}

Work
supported by the NTUA research program PEVE07.
G.~S.~was
supported in part by the US Department of Energy under grant
DE-FG05-91ER40627.

\end{document}